\documentclass[lettersize,journal]{IEEEtran}
\usepackage{amsmath,amsfonts}
\usepackage{graphicx}
\usepackage{algorithmic}
\usepackage{algorithm}
\usepackage{array}
\usepackage[caption=false,font=normalsize,labelfont=sf,textfont=sf]{subfig}
\usepackage{textcomp}
\usepackage{stfloats}
\usepackage{url}
\usepackage{verbatim}
\usepackage{graphicx}
\usepackage{cite}
\usepackage{color}
\usepackage{xcolor}
\usepackage{multirow}
\usepackage{multirow}
\usepackage{stfloats}
\usepackage{booktabs}
\usepackage{makecell}
\usepackage{threeparttable}

\begin{document}
 
\title{Binary Weight Multi-Bit Activation Quantization for Compute-in-Memory CNN Accelerators}

\author{Wenyong Zhou, Zhengwu Liu*, Yuan Ren, and Ngai Wong*
\thanks{This research was partially conducted by ACCESS – AI Chip Center for Emerging Smart Systems, supported by the InnoHK initiative of the Innovation and Technology Commission of the Hong Kong Special Administrative Region Government, and partially supported by the Theme-based Research Scheme (TRS) project T45-701/22-R of the Research Grants Council (RGC), Hong Kong SAR, and the National Natural Science Foundation of China Project 62404187. All authors are with the Department of Electrical and Electronic Engineering, The University of Hong Kong. *Corresponding authors: Zhengwu Liu, Ngai Wong.
}} 

\markboth{IEEE Transactions on Computer-Aided Design of Integrated Circuits and Systems,~Vol.~XX, No.~X, XXX~2025}%
{How to Use the IEEEtran \LaTeX \ Templates}

\maketitle
\begin{abstract}
Compute-in-memory (CIM) accelerators have emerged as a promising way for enhancing the energy efficiency of convolutional neural networks (CNNs). Deploying CNNs on CIM platforms generally requires quantization of network weights and activations to meet hardware constraints. However, existing approaches either prioritize hardware efficiency with binary weight and activation quantization at the cost of accuracy, or utilize multi-bit weights and activations for greater accuracy but limited efficiency. In this paper, we introduce a novel binary weight multi-bit activation (BWMA) method for CNNs on CIM-based accelerators. Our contributions include: deriving closed-form solutions for weight quantization in each layer, significantly improving the representational capabilities of binarized weights; and developing a differentiable function for activation quantization, approximating the ideal multi-bit function while bypassing the extensive search for optimal settings. Through comprehensive experiments on CIFAR-10 and ImageNet datasets, we show that BWMA achieves notable accuracy improvements over existing methods, registering gains of 1.44\%-5.46\% and 0.35\%-5.37\% on respective datasets. Moreover, hardware simulation results indicate that 4-bit activation quantization strikes the optimal balance between hardware cost and model performance.
\end{abstract}


\section{Introduction}
\label{sec:introduction}
Convolutional Neural Networks (CNNs) are pivotal in computer vision tasks, yet their growing complexity necessitates more computational power and energy~\cite{resnet,vgg}. Conventional digital circuits for CNN inference, such as central processing units (CPUs) and graphics processing units (GPUs), struggle with data movement inefficiencies due to their von Neumann architecture with separate memory and processing units~\cite{isaac,imce}. The compute-in-memory (CIM) architecture, which integrates processing and memory, offers a solution by largely reducing data movement~\cite{date,edtm}. Utilizing static random-access memory (SRAM), resistive random-access memory (RRAM), or ferroelectric field-effect transistor (FeFET), CIM-based accelerators enhance energy efficiency during CNN's most demanding task, namely, matrix-vector multiplications (MVMs), by executing in the analog domain within crossbar arrays and leveraging analog-to-digital converters (ADCs) and digital-to-analog converters (DACs) for data conversion~\cite{neurosim}. 
\begin{figure}[!t]
\centering
\includegraphics[scale=0.32]{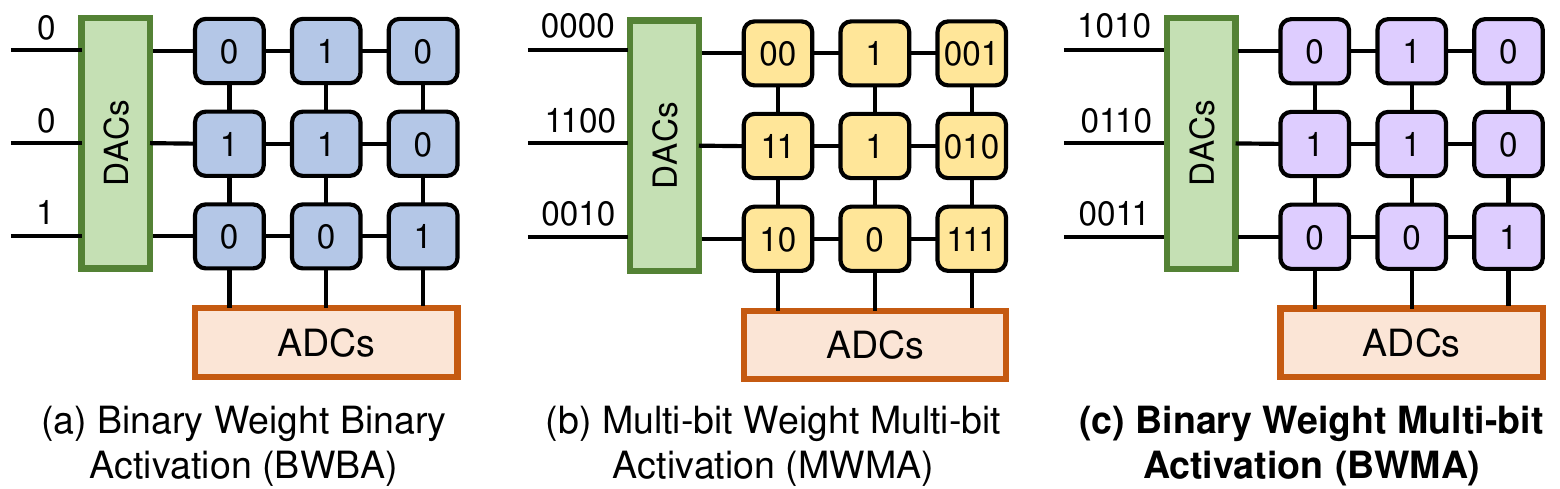}
\caption{Illustration of binary weight binary activation (BWBA, a), multi-bit weight multi-bit activation (MWMA, b), and our proposed binary weight multi-bit activation (BWMA, c) quantization on CIM-based accelerators.
}
\label{fig:motivation}
\vspace{-0.2cm}
\end{figure}

While CIM architectures offer promising solutions for CNN acceleration, they face practical implementation challenges due to hardware constraints and analog computing non-idealities. Model quantization, which reduces the precision of weights and activations, emerges as a crucial technique to address these challenges~\cite{reactnet,tang2017train}. However, quantization for CIM architectures requires special consideration. In CIM designs, weight precision directly impacts the area and power consumption of memory cells, while activation precision determines ADC/DAC complexity and energy cost~\cite{isaac}.

CIM-oriented quantization approaches have been developed utilizing two main strategies~\cite{huangetal, qpim, egq, sun2018xnor}. The first strategy, such as CIM-BNN~\cite{gu2023cim} and XNOR-RRAM~\cite{sun2018xnor}, focuses on extremely quantized networks with binary weights and binary activations (BWBA, see Fig.~\ref{fig:motivation}(a)) for efficient hardware implementations by replacing multiply-and-accumulate operations with XNOR and bit counting operations~\cite{sun2018xnor}. However, the strategy results in significant accuracy loss due to the radical quantization. The second strategy, such as CIMQ~\cite{cimq}, employs multi-bit weights and multi-bit activations (MWMA, see Fig.~\ref{fig:motivation}(b)) to maintain higher accuracy. Nevertheless, MWMA approaches face practical limitations: multi-bit weights increase storage overhead in CIM devices, while higher activation bitwidth impacts ADC costs and peripheral circuitry. Consequently, achieving both high accuracy and hardware efficiency remains challenging in CIM-oriented quantized networks.

To address these challenges, we propose a novel binary weight multi-bit activation (BWMA) quantization approach for CIM-based accelerators, as illustrated in Fig.~\ref{fig:motivation}(c). The contributions of this paper are multifaceted:
\begin{itemize}
    \item Unlike prior hardware-agnostic approaches, we propose a quantization framework that considers CIM's mixed-signal constraints by optimizing bitwidth based on cell precision and data converter resolution.
    
    \item We enhance model representation through two key innovations: a closed-form layer-specific weight binarization method and an efficient differentiable function for uniform multi-bit quantization, eliminating the need for exhaustive parameter search.
    
    \item Extensive evaluations validate our approach: achieving 0.35-5.46\% accuracy improvements on CIFAR-10 and ImageNet datasets, while hardware simulations across various CNN architectures reveal 4-bit data converters as the optimal trade-off between hardware cost and model performance.
\end{itemize}
\section{Related Work}
\label{sec:related}
Model quantization has emerged as a crucial technique for deep neural network deployment, with research spanning from binary to multi-bit precision schemes. Early binary neural networks demonstrated promising hardware efficiency, with FINN~\cite{finn} providing a scalable framework for BNN inference.
To address the accuracy challenges of binary networks, Tang et al.~\cite{tang2017train} introduced specialized training methods for compact binary networks, while ReactNet~\cite{reactnet} developed generalized activation functions to enhance BNN expressivity. Martinez et al.~\cite{martinez2020training} bridged full-precision and binary network training through real-to-binary convolutions, and balanced binary neural networks~\cite{shen2020balanced} improved information flow using gated residual mechanisms. Beyond binary quantization, PACT~\cite{pact} proposed parameterized clipping activations for flexible multi-bit quantization, enabling learnable quantization ranges that better preserve network accuracy. However, these quantization approaches primarily target conventional digital hardware, leaving room for specialized solutions for CIM architectures with unique characteristics and constraints.

There are many CIM-based CNN implementations. For example, ISAAC~\cite{isaac} explores an in-situ processing approach, where memristor crossbar arrays not only store input weights but also perform dot-product operations in the analog domain. IMCE~\cite{imce} employs parallel computational memory sub-arrays as fundamental units for bit-wise in-memory convolution operations. While both ISAAC and IMCE focus on hardware architecture design, our work emphasizes algorithm-level optimization and provides comprehensive evaluation to bridge the gap between network quantization and practical CIM implementation.
\section{Methodology}
\label{sec:methodology}
\subsection{Quantization on CIM Accelerators}
Due to design complexity and reliability concerns in various CIM devices (e.g., SRAM, RRAM, FeFET), we assume each cell in CIM crossbar arrays stores a 1-bit value, though different technologies demonstrate varied bit-width capabilities. While traditional BNNs~\cite{reactnet,tang2017train} constrain weights to fixed binary values ($\pm$1), limiting model representation across diverse CNN layers, CIM-based accelerators enable adaptive binary sets through layer-specific scaling factors. The weight-to-conductance mapping in CIM can be expressed as:
\begin{align}
G_{i,j} = \frac{w_{i,j}}{S_w} \frac{g_{max} - g_{min}}{\eta_w}
\label{eqn:map_1}
\end{align}
where $g_{max}$ and $g_{min}$ are conductance bounds, and scaling factors $S_w$ and $\eta_w$ align weight and conductance ranges. This mapping enables expanded weight representation through layer-specific binary values while maintaining hardware simplicity. The mixed-signal nature of CIM architectures necessitates multi-bit ADCs to convert analog matrix-vector multiplication outputs for digital processing. This hardware characteristic motivates our BWMA strategy - employing binary weights for efficient crossbar operations while maintaining higher precision in activations to match ADC resolution. Our BWMA framework jointly optimizes both components: it determines layer-specific binary weights by preserving statistical distributions (mean and standard deviation), while aligning activation quantization with ADC characteristics through a piecewise differentiable approximation.
\subsection{BWMA Quantization}
Our BWMA framework employs quantization-aware training (QAT) to achieve optimal performance. During training, we maintain both full-precision and quantized representations of weights and activations. The binary weight values are determined through moment matching, while activations are quantized to multiple bits using our proposed differentiable approximation function.

\begin{figure}[!t]
\centering
\includegraphics[scale=0.32]{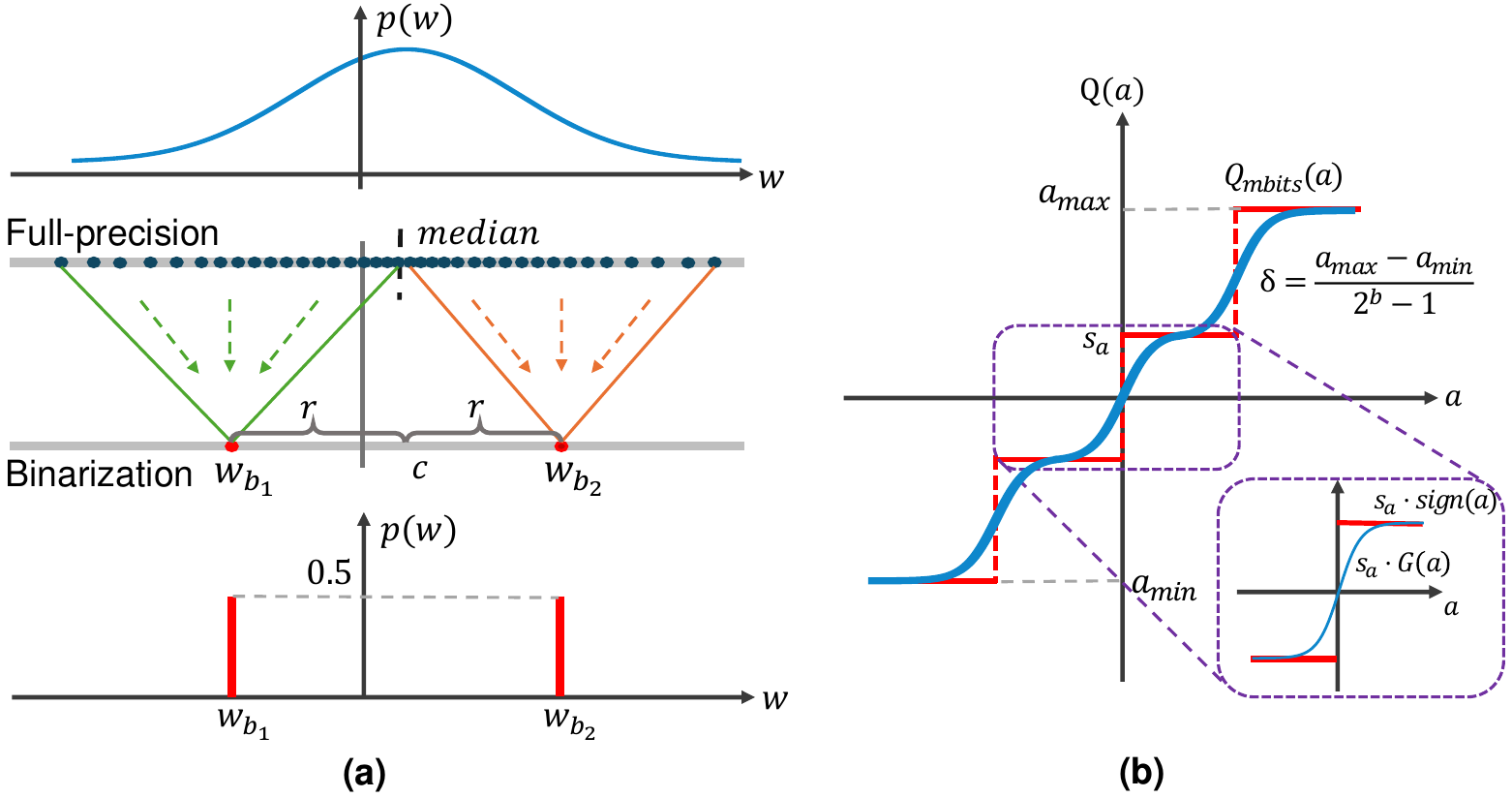}
\caption{BWMA quantization framework. Weight binarization (a) aligns the distributions' first and second moments (i.e., mean and standard deviation) before and after binarization. Multi-bit activation quantization (b) develops a novel differentiable function, consisting of scaled and shifted $G(a)$, to approximate the multi-bit uniform quantization.}
\label{fig:framework}
\vspace{-0.2cm}
\end{figure}

For the model weights $\{w_i\}_{i=1}^N$ in each layer, weight binarization maps them to two distinct values $w_{b_1}$ and $w_{b_2}$, as shown in Fig.~\ref{fig:framework}(a). Minimizing Kullback–Leibler (KL) divergence for binarization is intuitive but can be suboptimal due to calculation difficulties with uncertain priors~\cite{adabin}. Here, we propose aligning the first and second moments of distributions as an alternative to KL divergence for weight binarization. Considering that weights in CNNs generally follow a symmetric distribution~\cite{adabin}, we partition the weights through the median and evenly separate the weights as shown in Fig.~\ref{fig:framework}(a). By mapping the left half to $w_{b_1}$ and the right half $w_{b_2}$, our method is formulated as:
\begin{align}
\mathbb{E}[w_i]  = \frac{w_{b_1}+w_{b_2}}{2}
\label{eqn:1stmoment}
\end{align}
\begin{align}
\mathbb{E}[(w_i-\mathbb{E}[w_i])^2] = \frac{(w_{b_1}-\mathbb{E}[w_i])^2+(w_{b_2}-\mathbb{E}[w_i])^2}{2}
\label{eqn:2ndmoment}
\end{align}
where $\mathbb{E}[x]$ calculates the expected value (i.e. mean) of $x$. For ease of computation, we rewrite the binary values as $w_{b_1} = c - r$ and $w_{b_2} = c + r$, where $c$ is the midpoint between $w_{b_1}$ and $w_{b_2}$, and $r$ is a positive number representing the distance between $w_{b_i}$ ($i=1,2$) and $c$.
\begin{align}
c = \frac{1}{N} \sum_{i=1}^{N}w_i, \quad r =\sqrt{\frac{1}{N} \sum_{i=1}^{N} (w_i-c)^2}
\label{eqn:solution}
\end{align}
The terms on the right-hand side of Eq.~\ref{eqn:solution} are the mean ($\mu$) and standard deviation ($\sigma$) of original full-precision weights, respectively. 

Our BWMA framework achieves distribution alignment through moment matching. Specifically, for each layer during the forward pass, we compute the mean $\mu$ and standard deviation $\sigma_w$ of the original weights $\{w_i\}_{i=1}^N$. These statistics determine two binary values $w_{b_1} = \mu - \sigma_w$ and $w_{b_2} = \mu + \sigma_w$. To handle the non-differentiable binarization operation in the backward pass, we propose a modified straight-through estimator (STE):
\begin{equation}
\frac{\partial \mathcal{L}}{\partial w} = \frac{\partial \mathcal{L}}{\partial w_{out}} \cdot \begin{cases} 
\alpha \cdot (1-tanh^2(w\cdot t)) & \text{if } |w| \leq 1 \\
0 & \text{otherwise}
\end{cases}
\end{equation}
where $t$ is a temperature parameter that controls the steepness of gradient approximation, and $\alpha$ is a scaling factor.

Unlike the closed-form solution available for weight binarization, finding an analytical solution for activation quantization is infeasible due to the complex and asymmetrical distributions of activation values. This makes moment alignment ineffective in multi-bit scenarios. Consider the multi-bit uniform quantization $Q_{mbits}$ defined as follows:
\begin{align}
Q_{mbits}(a) = \mbox{round}(\frac{a}{\delta}) \cdot \delta, \quad \delta = \frac{a_{max}-a_{min}}{2^b-1}
\label{eqn:uniquant}
\end{align}
where $a_{min}$ and $a_{max}$ are the minimum and maximum values of $a \in [a_{min}, a_{max}]$, respectively, and $\delta$ represents the length of $2^b-1$ uniform intervals. The $\mbox{round}(\cdot)$ function returns the nearest integer. The non-differentiable $Q_{mbits}$ complicates the update of activations during backpropagation.

An empirical solution to this problem involves using a STE, which comprises differentiable functions~\cite{bnn, birealnet}. The $Q_{mbits}$ function, consisting of scaled and shifted $sign$ functions as illustrated in Fig.~\ref{fig:framework}(b), is approximated by a novel differentiable function aimed at simulating the $sign$ function. Historically, the derivative of $sign$ function, aka Dirac function, is estimated using either a rectangular~\cite{bnn} or triangular~\cite{birealnet} function (both computationally efficient but imprecise), or a piecewise function~\cite{adabin} which offers greater accuracy but is computationally burdensome. To overcome these, we introduce a quadratic function $g(a)$ as a differentiable approximation of the Dirac function:
\begin{equation}
g(a)=\begin{cases} 0, & |a| > 1 \\ -2a^2+\frac{5}{3}, & |a| \leq 1 \end{cases}
\end{equation}
The quadratic form of $g(a)$, akin to a bell-shaped curve, provides a superior approximation compared to constant or linear approaches, while also being computationally more efficient than complex piecewise functions. Integration of $g(a)$ results in $G(a)$ that approximates the $sign$ function:
\begin{equation}
G(a)=\left\{
\begin{aligned}
& -1, \quad &a < -1 \\
& -\frac{2}{3}a^3+\frac{5}{3}a, \quad &-1 \leq a \leq 1 \\
& 1, \quad &a > 1
\end{aligned}
\right.
\end{equation}
By scaling and shifting $G(a)$, namely, $s_aG(a-c_i)$ where $s_a$ and $c_i$ denote the scale and the center of each interval, respectively, we approximate the $Q_{mbits}$ function effectively.

\section{Experiments}
\label{sec:experiments}
\subsection{Experiment Setup}
\begin{figure}[!t]
\centering
\includegraphics[scale=0.28]{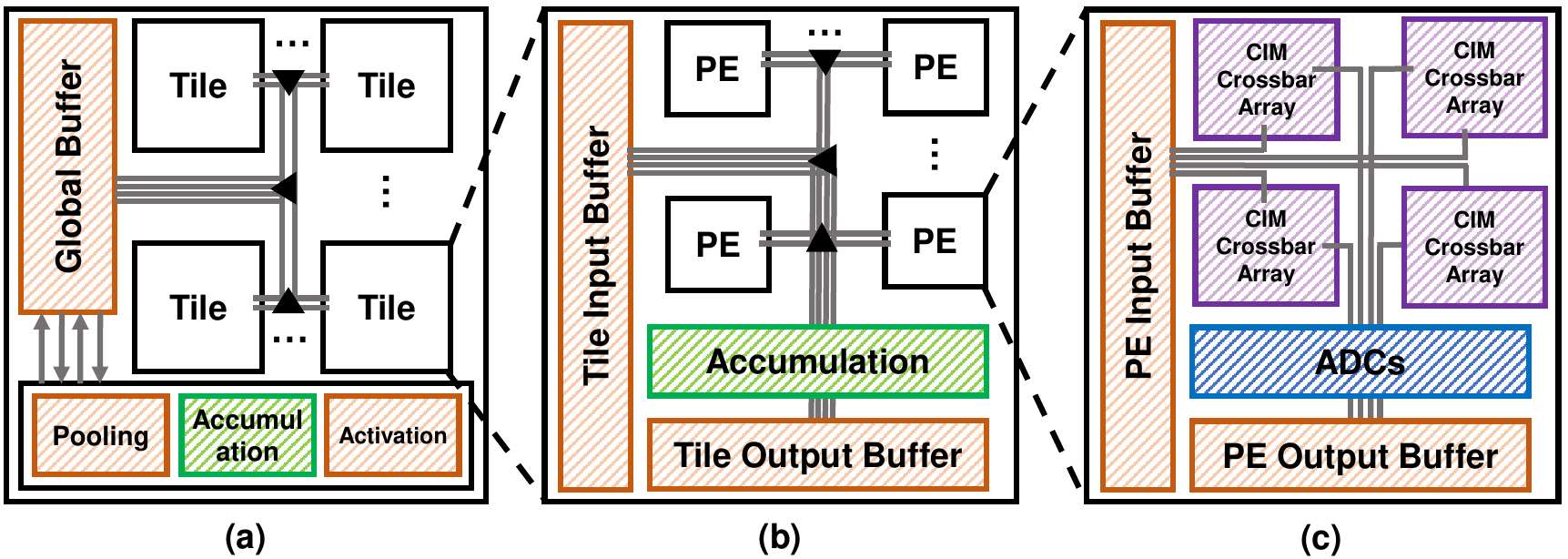}
\caption{Hierarchical architecture in CIM-based accelerators at chip (a), tile (b), and processing element (PE, c) levels. Components with different colors correspond to those parts in the figures below.}
\label{fig:system}
\vspace{-0.2cm}
\end{figure}

All experiments are conducted on NVIDIA RTX 3090 GPUs with 24 GB VRAM. The hardware simulations are fully dependent on the DNN+NeuroSim platform~\cite{neurosim}, which supports VGG and ResNet models across SRAM, RRAM, and FeFET devices. As shown in Fig.~\ref{fig:system}, DNN+NeuroSim leverages an H-tree routing architecture to efficiently manage data movement across different hierarchical levels. The routing infrastructure spans from chip-level interconnects down to individual PE arrays, facilitating communication between computational tiles, global buffers, and functional units. Moreover, the framework incorporates an optimized spatial mapping scheme for convolutional layers that partitions kernels based on their spatial locations into $K\times K$ sub-matrices, reducing buffer access requirements and improving data reuse efficiency. For linear layers, the framework employs a conventional mapping approach by unrolling weights into column vectors for matrix multiplication operations. 

\subsection{Experiment Results}
Table~\ref{tab:accuracy} compares the accuracy of ResNet-18 on CIFAR-10 and ImageNet datasets with our method with existing quantization schemes. The CIM-oriented MWMA methods quantize different parts of models, including weights~\cite{qpim}, activations~\cite{egq}, or both. The mixed precision of model parameters leads to the non-integers when calculating the average bitwidth of weights and activations. The CIM-oriented BWBA method~\cite{sun2018xnor} emphasizes their hardware implementation on RRAM-based CIM-based accelerators. Compared to previous CIM-oriented quantization methods, our BWMA approach with binary weights and 4-bit activation, an optimal balance between hardware cost and accuracy (explained later), improves the accuracy by 1.44\%-5.46\% and 0.35\%-5.37\% on the CIFAR-10 and ImageNet datasets, respectively.

Table~\ref{tab:binary_structures} compares our method with the vanilla counterparts of BDenseNet~\cite{BDenseNet}, MeliusNet~\cite{MeliusNet}, and ReActNet~\cite{reactnet}. Since these models are binary-specific structures, we set the activation to 1-bit. For BDenseNet28, our method achieves a 0.9\% improvement under the same training settings with negligible computational overhead. Similarly, when applied to MeliusNet and ReActNet-A structures, our method brings consistent performance gains.

\begin{figure}[!t]
\centering
\includegraphics[scale=0.38]{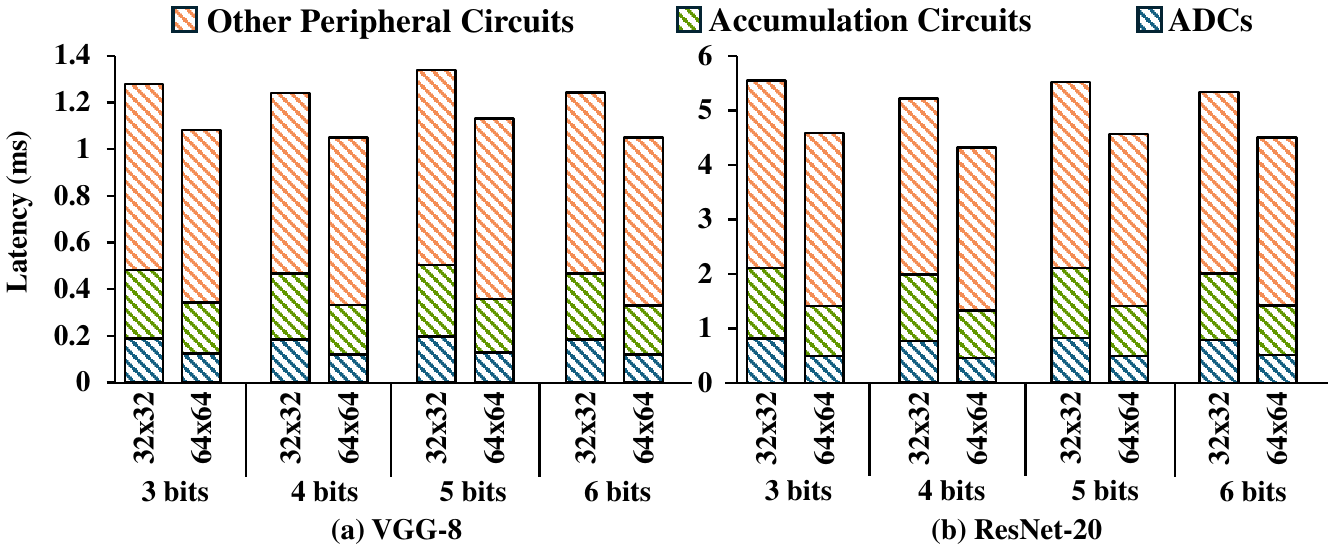}
\caption{The latency of VGG-8 (a) and ResNet-20 (b) on 32$\times$32 and 64$\times$64 RRAM crossbars.}
\label{fig:hwcost_la}
\vspace{-0.2cm}
\end{figure}

\begin{figure}[!t]
\centering
\includegraphics[scale=0.3]{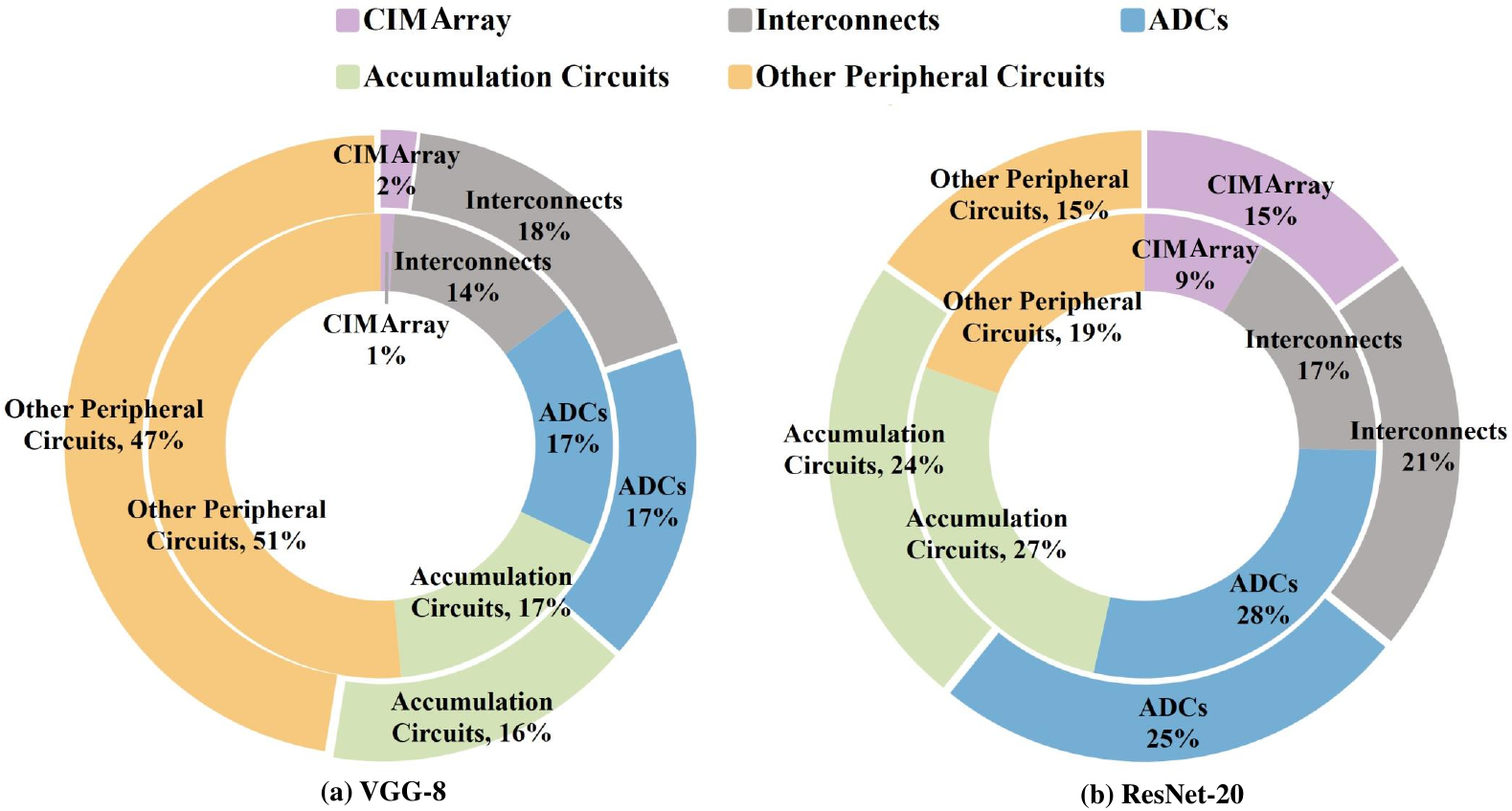}
\caption{The chip area of VGG-8 (a) and ResNet-20 (b) on 32$\times$32 (inner ring) and 64$\times$64 (outer ring) RRAM crossbars.}
\label{fig:hwcost_area}
\vspace{-0.2cm}
\end{figure}

\begin{figure}[!t]
\centering
\includegraphics[scale=0.35]{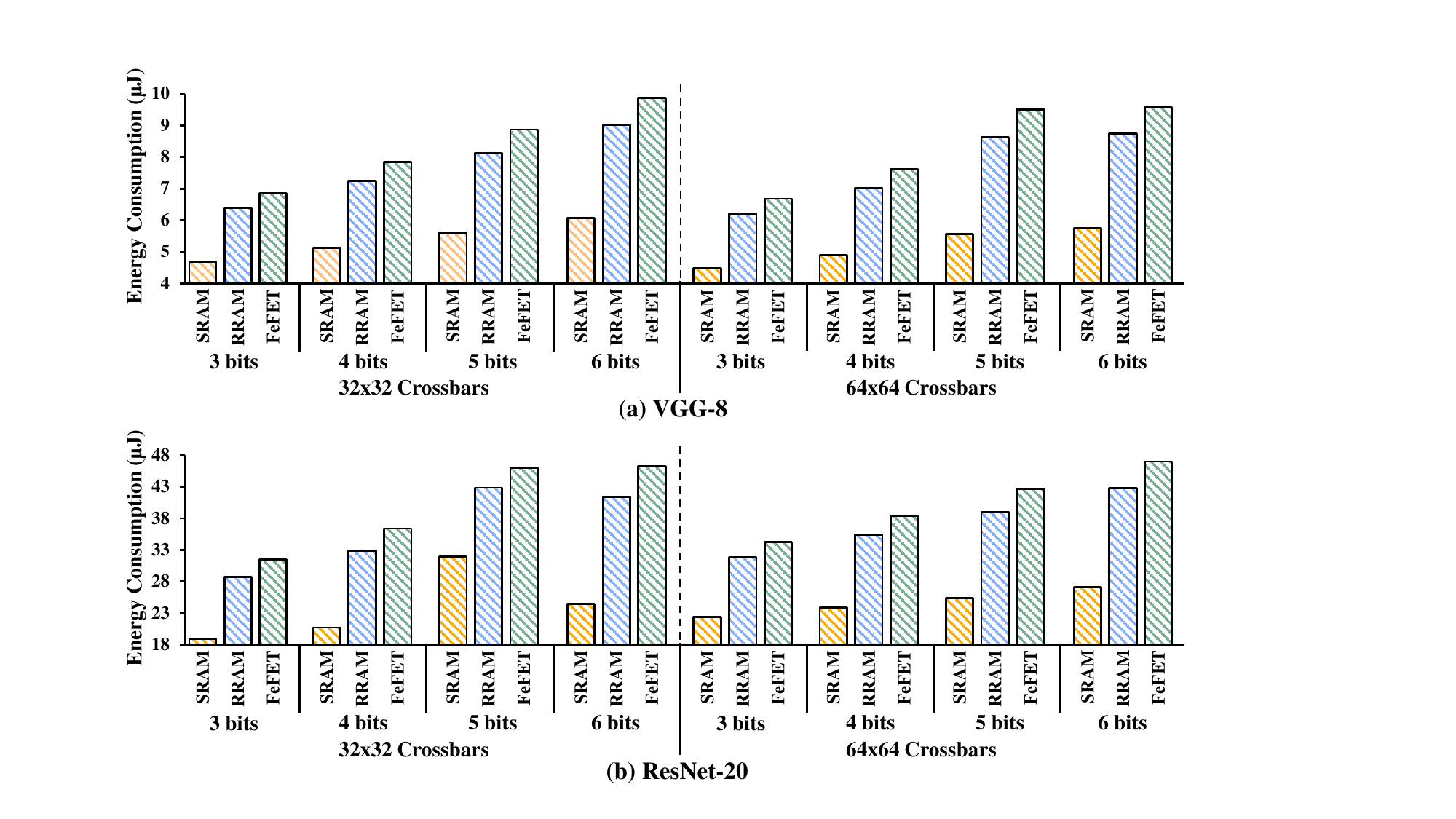}
\caption{The energy consumption of VGG-8 (a) and ResNet-20 (b) on different crossbars, including different crossbar sizes and device types.}
\label{fig:hwcost_eng}
\vspace{-0.2cm}
\end{figure}

For hardware metrics, we simulate latency, chip area, and energy consumption of VGG-8 and ResNet-20 on the CIFAR-10 dataset. The performance are reported in Figures~\ref{fig:hwcost_la}, \ref{fig:hwcost_area} \&~\ref{fig:hwcost_eng}. Fig.~\ref{fig:hwcost_la} depicts the total latency and breakdown of different components of VGG-8 and ResNet-20 on different crossbar sizes, taking RRAM-based accelerators as an example. According to the simulation results, the ADCs contribute roughly 14\% in total latency, while the accumulation circuits (at PE and tile levels) and other peripheral circuits are the primary consumers (accounting for 23\% and 63\%). Although the absolute value significantly varies for different CNNs, the relative relationship between multiple components is stable, indicating the similarity of CIM-based accelerators for accelerating CNNs. Increasing crossbar size brings lower latency, which coincides with the trend in Fig.~\ref{fig:hwcost_eng}, demonstrating that larger crossbars could improve hardware efficiency.
\begin{table}[!t]
\renewcommand{\arraystretch}{1.2}
\caption{Accuracy of ResNet-18 quantization with existing methods and our method on CIFAR-10 and ImageNet datasets.}
\centering
\resizebox{\columnwidth}{!}{
\begin{tabular}{lccc}
\toprule
\multirow{2}{*}{\textbf{Methods}} & \multicolumn{2}{c}{\textbf{Bitwidth}} & \multirow{2}{*}{\textbf{Accuracy}} \\
\cmidrule(lr){2-3}
& \textbf{Weight} & \textbf{Activation} & \\
\midrule
Huang \textit{et al.}~\cite{huangetal} & 6.1 & 6.3 & 84.45\% / 62.48\% \\
Q-PIM~\cite{qpim} & 5.2 & 8.0 & 86.23\% / 63.83\% \\
EGQ~\cite{egq} & 6.1 & 6.3 & 87.72\% / 67.50\% \\
XNOR-RRAM~\cite{sun2018xnor} & 1 & 1 & 88.47\% / N/A \\
\textbf{BWMA (Ours)} & \textbf{1} & \textbf{4} & \textbf{89.91\% / 67.85\%} \\
\bottomrule
\end{tabular}
}
\label{tab:accuracy}
\vspace{-0.2cm}
\end{table}
\begin{table}[!t]
\renewcommand{\arraystretch}{1.2}
\caption{Comparisons on ImageNet for binary-specific structures.}
\centering
\resizebox{\columnwidth}{!}{
\begin{tabular}{llcc}
\toprule
Networks & Methods & OPs ($\times 10^8$) & Top-1 (\%) \\
\midrule
\multirow{2}{*}{BDenseNet28~\cite{BDenseNet}} & Origin & 2.09 & 62.6 \\
& \textbf{BWMA} & 2.11 & 63.5 (+0.9) \\
\multirow{2}{*}{MeliusNet22~\cite{MeliusNet}} & Origin & 2.08 & 63.6 \\
& \textbf{BWMA} & 2.10 & 64.3 (+0.7) \\
\multirow{2}{*}{MeliusNet29~\cite{MeliusNet}} & Origin & 2.14 & 65.8 \\
& \textbf{BWMA} & 2.17 & 66.1 (+0.3) \\
\multirow{2}{*}{ReActNet-A~\cite{reactnet}} & Origin & 0.87 & 69.4 \\
& \textbf{BWMA} & 0.88 & 70.0 (+0.6) \\
\bottomrule
\end{tabular}
}
\label{tab:binary_structures}
\vspace{-0.2cm}
\end{table}

Fig.~\ref{fig:hwcost_area} compares the impact of crossbar size and model architecture on RRAM-based accelerators regarding chip area. It is clear that the CIM array only takes a small amount in the chip area, ranging from 1\% to 15\%. On the one hand, those cells in crossbar arrays in SRAM-based accelerators are much larger than those in RRAM/FeFET-based accelerators. Unlike SRAM, which comprises 6 transistors in one cell, the typical configuration in an RRAM cell is one transistor and one RRAM device (1T1R). On the other hand, other peripheral circuits take a smaller portion in SRAM-based accelerators than in RRAM/FeFET-based accelerators, making the total chip areas differ slightly across different device types. Together with a negligible contribution of CIM arrays on latency and energy consumption, optimizing the data conversion process and peripheral circuits play a vital role in reducing the hardware cost of CIM-based accelerators.

Fig.~\ref{fig:hwcost_eng} compares the impact of crossbar size and device type on energy consumption in various quantization scenarios. From the figure, SRAM-based accelerators consume 57\% to 73\% of energy than emerging devices due to the lower working voltage. The higher working voltage of FeFET-based accelerators leads to more energy consumption than RRAM-based accelerators. Although larger crossbars are typically more hardware efficient, the expected descending trend is not observed in both VGG-8 and ResNet-20, mainly due to the occupation of the unused cells increased (from 6.5\% to 30.4\% and from 10.7\% to 39.1\% for VGG-8 and ResNet-20, respectively) when opting for larger crossbars. Therefore, maintaining a high resource utilization should be a key concern in determining suitable crossbar size.

\begin{table}[!t]
\renewcommand{\arraystretch}{1.2}
\caption{Performance comparison of Mamba and Mamba2 models under noise.}
\centering
\resizebox{\columnwidth}{!}{
\begin{tabular}{>{\centering\arraybackslash}p{1cm}cccc}
\toprule

\textbf{Models} & \textbf{Size} & \textbf{Clean} & \textbf{Baseline} & \textbf{Ours}   \\ \hline
\multirow{4}{*}{Mamba}           & 3 bits                          & 86.21\%        & 1.00 / 1.00       & 1.00 / 1.00     \\
                                 & 4 bits                          & 87.06\%        & 1.16 / 1.19       & 1.14 / 1.19     \\
                                 & 5 bits                          & 87.15\%        & 1.32 / 1.39       & 1.28 / 1.37     \\
                                 & 6 bits                          & 87.37\%        & 1.48 / 1.59       & 1.42 / 1.57     \\ \cmidrule(lr){1-5}
\multirow{4}{*}{Mamba2}          & 3 bits                          & 86.53\%        & 1.00 / 1.00       & 1.00 / 1.00     \\
                                 & 4 bits                          & 87.35\%        & 1.16 / 1.20       & 1.14 / 1.19     \\
                                 & 5 bits                          & 87.59\%        & 1.32 / 1.39       & 1.28 / 1.38     \\
                                 & 6 bits                          & 87.68\%        & 1.48 / 1.59       & 1.43 / 1.57     \\
\bottomrule
\end{tabular}
}
\label{tab:mamba_performance}
\end{table}

Our multi-bit activation quantization enables flexible accuracy-hardware trade-offs. To quantify these trade-offs, we analyze hardware costs across various data converter resolutions (Table~\ref{tab:avghwcost}), normalizing latency, area, and energy metrics against 3-bit converters. Although SRAM- and RRAM-based implementations exhibit distinct absolute performance characteristics, their normalized costs follow similar trends with increasing resolution. Higher-resolution converters (5-/6-bit) introduce substantial hardware overhead without commensurate accuracy gains, while 4-bit precision consistently emerges as the optimal balance point across all device types, architectures, and crossbar sizes. This finding demonstrates that our BWMA framework not only enhances model accuracy but also provides clear guidelines for hardware-efficient CIM accelerator design.
\section{Conclusion}
\label{sec:conclusion}
This paper has proposed a hardware-aware quantization framework for CIM-based accelerators that optimizes both cell precision and data converter resolution. Our key innovations include analytically-derived layer-specific weight binarization through moment matching, and an efficient differentiable approximation for uniform multi-bit quantization. Experiments demonstrate superior accuracy with 1.44\%-5.46\% and 0.35\%-5.37\% improvements on CIFAR-10 and ImageNet respectively, while hardware simulations across SRAM, RRAM, and FeFET implementations identify 4-bit data converters as the optimal balance between cost and performance.


\small
\bibliographystyle{ieeetr}

\bibliography{bare_conf}

\vfill

\end{document}